\input{psfig.sty}

\def\selectedoptions{final}

\documentclass[
   \selectedoptions
  ]
  {aipproc}

\def\selectedlayoutstyle{6x9}
\layoutstyle\selectedlayoutstyle

\SetInternalRegister\hbadness{8000} 

\newcommand\doingARLO[2][]{%
  \ifx\mmref\undefined #1\else #2\fi
}
  
\newcommand{\etal}{{\it et al. }}
\begin{document}

\title 
      []
      {Relativistic Effects on X-ray Emissions from Accretion Disks 
around Black Holes}


\author{Xiaoling Zhang}{
  address={University of Alabama in Huntsville and National Space Science and Technology Center, \\Physics Department, Huntsville, AL 35899, USA},
  email={xizhang@jet.uah.edu},
  homepage={http://jet.uah.edu/~xizhang}
}

\iftrue
\author{Shuang Nan Zhang}{
  address={University of Alabama in Huntsville and National Space Science and Technology Center, \\Physics Department, Huntsville, AL 35899, USA},
  email={zhangsn@jet.uah.edu},
  homepage={http://jet.uah.edu/~zhangsn}
}
\author{Yangsen Yao}{
  address={University of Alabama in Huntsville and National Space Science and Technology Center, \\Physics Department, Huntsville, AL 35899, USA},
  email={yaoys@jet.uah.edu},
  homepage={http://jet.uah.edu/~yaoys}
}

\fi

\copyrightyear  {2001}

\begin{abstract}
Special and general relativistic effects on the X-ray emissions, 
especially the continuum spectra, from the accretion disks around black holes 
are investigated using the ray-tracing method. Because both the special and general relativistic 
effects are more important at distances closer to the black hole, 
the relativistic modifications to the emitted X-ray spectra are more significant 
at higher energies. In a simple accretion disk precession model, the relativistic effects 
can account for the energy dependence of the QPO amplitude and phase-lags observed in several black hole binaries. 
The dramatic QPO phase-lag transitions only show up 
when the central black hole is spinning rapidly. The narrow distribution of the observed 
system inclination angles of black hole binaries may be due to the selection effect caused by
the relativistic effects around black holes.
\end{abstract}

\date{\today}
\maketitle

\section{Introduction}
\vspace{-3mm}
Many black hole binaries have been known and studied in recent years.
Because of the strong gravitational field near the black hole, 
the radiation from the inner portion of the accretion disk experiences 
strong relativistic modifications before escaping from the system. 
In order to infer the physical parameters of the accretion disk and the black hole from the observed X-ray radiation, 
it is necessary to consider the relativistic effects (e.g., Zhang, Cui \& Chen 1997; Cui, Zhang \& Chen 1998).
The exact modifications depend strongly on the angular momentum of the black hole and 
the inclination angle of the disk, as well as the black hole mass and the accretion rate.

Analytical calculations of the relativistic modifications can only be done in a few very simple cases. 
For most applications, numerical methods must be adopted. 
Here we use the ray-tracing method by Fanton \etal  (1997), 
assuming that the disk is a thin Keplerian disk. 
The ray-tracing method takes all special and general relativistic effects into account.

In this paper we calculate the relativistically modified X-ray spectra from 
black hole X-ray binaries with different angular momenta and disk 
inclination angles. In particular we investigate the X-ray light curve 
modulations if the accretion disk is assumed to precess. Our calculations 
reproduce the observed light curve modulation  {\it rms}  (root-mean-squares) and 
the peculiar phase-lags around QPO peaks from the microquasar GRS1915+105, if the black hole in
GRS1915+105 is spinning rapidly. Our results thus suggest that the observed 
light curve modulations and phase-lags are manifestation of the strong 
relativistic effects around a rapidly spinning black hole. We also propose 
that the observed system inclination angles around 60 to 70 degrees 
for all black hole binaries may be due to the selection effect caused by
the relativistic effects around black holes.

\section{Models for X-ray Spectra and Flux Modulations}
\vspace{-3mm}
We follow the widely established model for the X-ray spectra from black 
hole X-ray binaries, i.e., the two component model consisting of a soft, blackbody-like 
component and a hard, power-law like component. The soft component is 
usually approximated by the {\it diskbb} model in the 
{\it XSPEC} package. Here we use the radial temperature profile derived in the Kerr metric for the soft component.
 
For the hard component, currently there is no widely accepted physical 
model yet. For simplicity, we assume that the hard component is also produced from 
very close to the accretion disk. We further assume that the local hard component is of a 
power-law shape with a low energy cutoff. The local power-law photon index 
is assumed to be
$\alpha = 2.2 + (r/r_g)/40$ in order to mimic the observed power-law shape. 
The low energy turn-over is the same as that in the Comptonization model (Sunyaev and Titarchuk 1980) 
i.e., implying that the power-law is produced via (thermal or non-thermal) Comptonization and 
the seed photons are the local blackbody emission 
from the disk. The emissivity of the hard and soft component is assumed to 
be the same, i.e., the gravitational energy 
release to X-ray radiation is shared equally in the soft and hard 
component.

We consider all relativistic effects when calculating the X-ray spectra at infinity. 
First, the rapid Keplerian motion of the disk causes both the Doppler frequency shift and Doppler boosting.  
Secondly, the strong gravitational field introduces the gravitational redshift and focusing.
These effects can be expressed with the transfer function (Cunningham, 
1975), calculated here with the ray-tracing method of Fanton \etal  (1997).

\section{Results of Ray-tracing calculations}
\vspace{-3mm}
In Fig. 1, we show the calculated X-ray spectra at infinity, for different 
black hole angular momenta and disk inclination angles. We fixed the black 
hole mass and mass accretion rate, such that the peak temperature for the 
extremal Kerr black hole is 2.1 keV; for all other black hole angular 
momenta, the corresponding peak temperatures are lower accordingly.
These spectra may be fitted using the XSPEC package with the RXTE PCA response and 
{\it diskbb+powerlaw} model, with reasonable parameters and residuals, 
demonstrating that our spectral model can indeed reproduce the 
observed X-ray spectra from black hole X-ray binaries. 

\begin{figure}
\label{spectrum}
\psfig{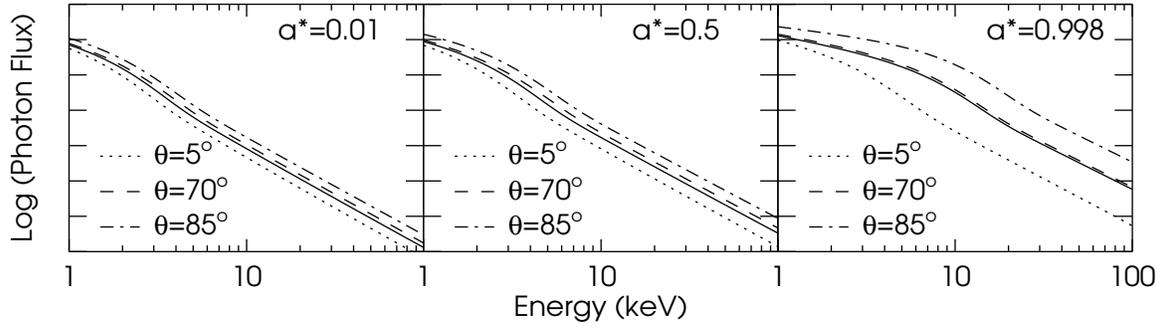}
\caption { Accretion disk spectra around black holes with different angular momenta and viewed with different
inclination angles from infinity. The solid lines are the local spectra and 
the dotted lines are the spectra observed at infinity, after modifications by special and general 
relativistic effects. For a rapidly spinning black hole system, the observed spectrum 
at infinity above 10 keV may deviate from the local spectrum significantly.
}
\end{figure}

\begin{figure}
\psfig{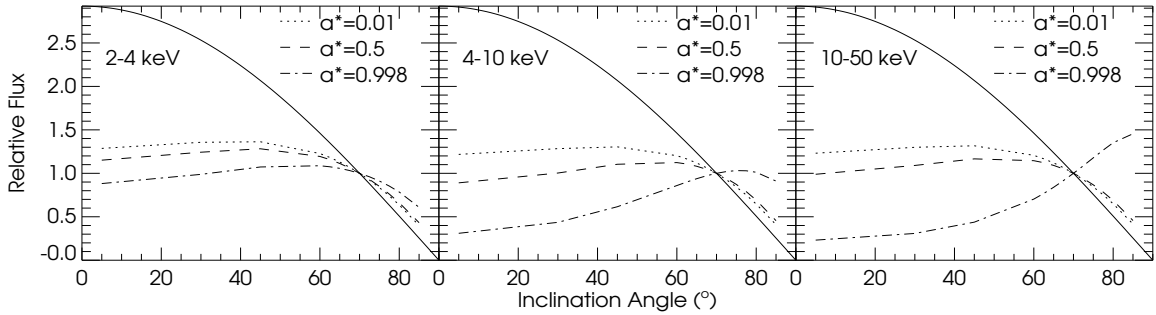}
\caption{Flux at infinity in different energy bands as a function of 
the inclination angle after the relativistic modifications, as shown by the dotted lines; 
the solid lines are for the cosine law, appropriate when 
relativistic effects are negligible. The deviations from the cosine law are very significant for a rapidly spinning
black hole, especially in high energy bands. This indicates that the observed light curve modulations in different energy
band will not be in phase when the disk is forced to precess.}
\end{figure}
\begin{figure}
\begin{minipage}{0.60\textwidth}
\psfig{figure=lightcurves.epsi,width=3.5in} 
\end{minipage}
\begin{minipage}{0.01\textwidth}
\hspace{0.001mm}
\end{minipage}
\begin{minipage}{0.35\textwidth}
{\bf \small FIGURE 3.} {\small Observed light curves in different energy bands at infinity. 
The whole accretion disk responsible for X-ray emission is assumed to 
precess at a certain frequency and the precession is assumed to be the only 
source of the observed light curve modulation. Note that for a non- and slowly spinning black hole, 
the low and high energy light curves are in phase. However when $a^*=0.998$, the light curves in different energy bands are
very different; low energy and high energy light curves are in fact opposite in phase.}
\end{minipage}
\label{lc}
\end{figure}

In Fig. 2, we show the observed flux in different energy bands as a 
function of the disk inclination angle. It is clear that significant 
deviations from the cosine law exist for a rapidly spinning black hole, 
especially at higher energies. This demonstrates clearly the necessity of taking into account relativistic
effects if the black hole is spinning. 

In a simple disk precession model, we may assume that the whole inner disk region precesses at a certain period, 
which is determined by the exact physical mechanism and mode of the disk precession.
The observed X-ray flux will then display periodic oscillations. 
As an example, we assume that the disk between the last stable orbit 
and 200 $r_g$ undergoes stable precession with inclination angles between 60 and 80 
degrees. No other source of flux modulation is assumed to exist. In Fig. 3, 
the observed light curves in different energy bands are plotted for different black hole 
angular momenta. It is striking that the low energy and high energy light 
curves are opposite in phase for a rapidly spinning black hole.

In Fig. 4, we show two sample power density spectra and the 
phase-lag spectra between them for different black hole angular momenta, with or without any white noise. Relatively strong 
harmonics of the precession frequency are observed. When $a^*=0.5$ no phase-lag is present, because the low and high energy 
light curves are in phase. However, when $a^*=0.998$, the low and high energy
light curves are always opposite in phase. However, in the existence of a weak white noise, 
only sharp phase-lag transitions are observed around the peaks of 
the power density spectra. 

Finally in Fig. 5, we show the  {\it rms} of the light curve modulations as a 
function of energy for different black hole angular momenta. For non- and slowly spinning black holes, 
the light curve modulation  {\it rms} does not change significantly as a function of energy. 
However, for the case of an extremal Kerr black hole, the light curve modulation depends 
strongly on the photon energy. 

In summary, when the black hole is spinning rapidly, sharp phase-lag 
transitions and strong energy dependence of the X-ray light curves are 
expected. Therefore the X-ray light curves may be used for 
probing the strong relativistic effects near black holes.
\setcounter{figure}{3}
\begin{figure}
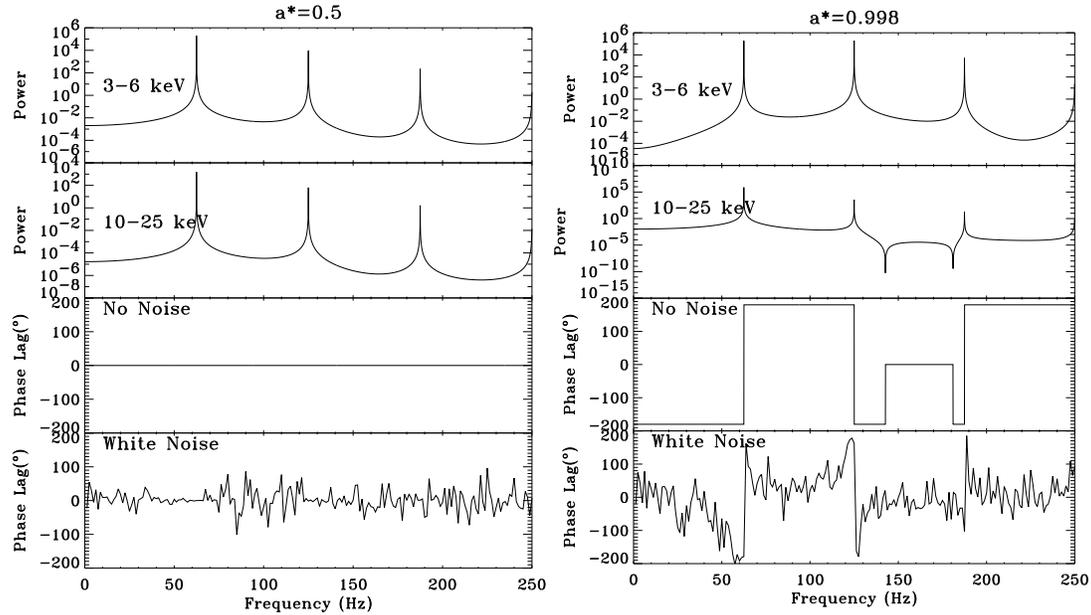

\begin{minipage}{0.50\textwidth}
\psfig{figure=3-6and10-25-1-crossspectrum.epsi,width=2.8in}
\end{minipage}
\begin{minipage}{0.50\textwidth}
\psfig{figure=3-6and10-25-2-crossspectrum.epsi,width=2.8in}
\end{minipage}
\caption{Sample power density and phase lag spectra, 
for two light curves in 3-6 keV and 10-25 keV bands. The white noise is 
about 0.2\% of the signals.
For the extremal Kerr black hole ($a^*=0.998$), 
sharp phase-lag changes are expected around the peaks of the power density
spectrum. The sharp phase-lag changes are purely due to the relativistic effects around a rapidly spinning black hole.}
\label{lag}
\end{figure}

\begin{figure}
\begin{minipage}{0.60\textwidth}
\psfig{figure=rms.epsi,width=3.3in}
\end{minipage}
\begin{minipage}{0.01\textwidth}
\hspace{0.01mm}
\end{minipage}
\begin{minipage}{0.35\textwidth}
{\bf \small FIGURE 5.} {\small The  {\it rms} (root-mean-squares) of the light curve modulations, as a 
function of energy.
For non- and slowly spinning black holes, 
the light curve modulation  {\it rms} does not change significantly as a function of energy. 
However, for the case of an extremal Kerr black hole, the light curve modulation depends 
strongly on the photon energy.}
\end{minipage}
\label{rms}
\end{figure}

\section{Discussions}
\vspace{-3mm}
The most striking result of this investigation is the naturally produced sharp 
phase-lag transitions around the peaks in the power-density spectra. Regardless the details of our
calculations, this feature is robust and relies essentially on only three assumptions: higher energy photons are produced at distances closer
to the black hole, the periodic flux modulations are mainly due to the accretion disk precession, and the black hole is spinning rapidly.
This phenomenon has been observed in GRS1915+105 (Cui, 1999; Lin \etal, 2000). 
The increasing trend of the light curve  {\it rms} for an extremal Kerr black hole in the middle 
energy range, as shown in Fig. 5,  has also been detected in GRS1915+105 (e.g., Cui, 1999). Therefore our results confirm that the black hole in GRS1915+105 
is spinning rapidly (Zhang, Cui \& Chen 1997; Cui, Zhang \& Chen 1998). From Fig. 5, 
we predict that the  {\it rms} decreases at energies above about 20-30 keV for extremal Kerr black hole 
binaries.

Finally our calculations may shed light 
on the outstanding puzzle that almost all observed X-ray black hole 
binaries have inclination angles around 60$^\circ$-70$^\circ$ (Zhang \etal  1997). 
The combined effects of higher space density for high inclination
angle systems and the relative flux of these high inclination angle systems (see Fig. 2) favor detection 
of binary systems at inclination angles around 60$^\circ$-70$^\circ$. Thus it is 
likely that the inclination angle distribution of known black hole binaries is simply a 
selection effect favoring the brightest X-ray sources, manifesting the strong relativistic effects around black holes.

{\bf Acknowledgments:} We thank Dr. Fanton for providing us with the ray-tracing code. 
We have also benefited from discussions with
Drs. Lev Titarchuk, Wei Cui, Yuxin Feng and Dingxiong Wang. This work was supported in part by NASA 
MSFC under contract NCC8-200 and by
NASA LTSA Program under grants NAG5-7927 and NAG5-8523. 


\end{document}